\documentstyle[preprint,aps,epsf,floats]{revtex}

\def\NPB{Nucl. Phys. B }
\def\PLB{Phys. Lett. B }
\def\PRL{Phys. Rev. Lett. }
\def\PRD{Phys. Rev. D }

\def\etal{{\it et.~al.}}

\def\be{\begin{equation}}
\def\ee{\end{equation}}
\def\bea{\begin{eqnarray}}
\def\eea{\end{eqnarray}}
\def\bean{\begin{eqnarray*}}
\def\eean{\end{eqnarray*}}
\def\bary{\begin{array}}
\def\eary{\end{array}}
\def\bi{\bibitem}
\def\bit{\begin{itemize}}
\def\eit{\end{itemize}}

\def\ne{\nu_e}
\def\nm{\nu_{\mu}}
\def\nt{\nu_{\tau}}

\begin{document}

\title{
\bf $CP$ Violation for Leptons at Higher Energy Scales}

\author{{Cheng-Wei Chiang}}

\address{Department of Physics, Carnegie Mellon University,
Pittsburgh, Pennsylvania 15213}

\maketitle

\begin{abstract}
The phase convention independent measure of $CP$ violation for three
generations of leptons is evaluated at different energy scales.
Unlike in the quark sector, this quantity does not vary much between
the weak and the grand unification scales.  The behavior of the
measure of $CP$ violation in the Standard Model is found to be
different from that in the extensions of the Standard Model.
\end{abstract}
\vspace{0.2in}

\newpage

with the discovery of neutrino mixing, which indicates nonzero
neutrino masses \cite{SK1998}, there recently has been a lot of
interest in neutrino physics, including measuring possible $CP$
violation in the leptonic sector of the three-generation Standard
Model \cite{BWP1980}.  $CP$ nonconservation in the leptonic sector has
important implication in leptogenesis in the early universe, as it can
feed into a baryon asymmetry through the sphaleron mechanism
\cite{FY1986}.  One of the necessary conditions to have baryogenesis
or leptogenesis is $CP$ violation \cite{S1967}.  It is thus natural to
discuss $CP$ nonconservation in the leptonic sector at higher energy
scales.

In analogy with the quark sector, the Maki-Nakagawa-Sakata (MNS)
matrix describes the mixing among the flavor eigenstates
$\nu_{\alpha}$ and mass eigenstates $\nu_i$ of neutrinos with
$\alpha=e,\mu,\tau$ and $i=1,2,3$ \cite{MNS1962}
%
%
\be
\label{MNS}
V_{MNS} = 
\left(
\bary{ccc}
c_{12}c_{13} & s_{12}c_{13} & s_{13}e^{-i\delta} \\
-c_{23}s_{12}-c_{12}s_{23}s_{13}e^{i\delta} & 
c_{12}c_{23}-s_{12}s_{23}s_{13}e^{i\delta} & c_{13}s_{23} \\
s_{12}s_{23}-c_{12}c_{23}s_{13} e^{i\delta} & 
-c_{12}s_{23}-c_{23}s_{12}s_{13}e^{i\delta} & c_{23}c_{13}
\eary
\right),
\ee
where $c_{ij} \equiv \cos \theta_{ij}$, $s_{ij} \equiv \sin
\theta_{ij}$ ($0 \leq \theta_{ij} \leq \pi/2$), and $\delta$ is a
phase.  Instead of referring to the above phase-dependent matrix, it
has been proven that with three generations of fermions $CP$ is not
conserved if and only if the single parametrization-independent
quantity
\be
\label{detC}
{\rm det}\,C = {\rm det} [N N^{\dagger},E E^{\dagger}]
\ee
is nonvanishing \cite{J1985}, where $N$ and $E$ denote the $3\times3$
Yukawa coupling matrices for the neutrinos and charged leptons,
respectively\footnote{As pointed out in Ref.~\cite{X2000}, the
definition in Eq.~(\ref{detC}) implies that the neutrinos are
Dirac-type particles.  If one has Majorana neutrinos, the definition
should be ${\rm det}\,C = {\rm det} [N^{\dagger} N,E E^{\dagger}]$
which, however, does not affect later analyses.}.  In terms of
$V_{MNS}$ matrix elements,
\be
\label{detCexp}
{\rm det}\,C = 
2 (m_{\tau}^2-m_{\mu}^2) (m_{\mu}^2-m_e^2) (m_e^2-m_{\tau}^2)
(m_{\nt}^2-m_{\nm}^2) (m_{\nm}^2-m_{\ne}^2) (m_{\ne}^2-m_{\nt}^2) J,
\ee
where $m_{\nt\nm\ne}$ and $m_{\tau,\mu,e}$ are eigenvalues of $N$ and
$E$, respectively, and $J=s_{12} s_{23} s_{13} c_{12} c_{23} c_{13}^2
\sin \delta$ is an invariant proposed by Jarlskog in the quark sector
\cite{J1985}.  The quantity ${\rm det}\,C$ vanishes if (i) any two
leptons with the same charge have the same mass; (ii) any of the
angles $\theta_{ij}$ has the value $0$ or $\pi/2$, or (iii) $\delta=0$
or $\pi$.\footnote{These conditions are not independent of one another;
see, for example, Ref.~\cite{FX1999}.}

To study the scale dependence of ${\rm det}\,C$, it is natural to
solve its renormalization group equation (RGE) given a set of initial
values for the relevant quantities such as the Yukawa and gauge
couplings at the weak scale.  Unlike the situation in the quark sector
where the analogous quantity drops by 4 to 5 orders of magnitude and
may provide insufficient $CP$ violation for baryogenesis
\cite{ADG1986}, we will show that this quantity for leptons does not
change very much between the weak and grand unification theory (GUT)
scales, which we take to be $\sim 10^{16}$ GeV.  Since $CP$ violation
is one of the ingredients for baryogenesis or leptogenesis
\cite{S1967}, this may have an important consequence for the evolution
of the early universe.

We will consider two methods for neutrinos to acquire mass: (1) the
neutrinos have the Dirac masses generated through the the same
mechanism as the other fermions; and (2) the neutrinos obtain Majorana
masses through the seesaw mechanism \cite{Y1979}.  We will study the
situation in the Standard Model (SM), the two Higgs doublet model
(THDM) and the minimal supersymmetric model (MSSM).

As shown in Ref.~\cite{ADG1986} for quarks, ${\rm det}\,C$ obeys a
rather simple equation under the renormalization group.  In the
leptonic sector, this equation can be derived from the renormalization
group equations for $N$ and $E$.  They can be written in the following
form:
\begin{mathletters}
\label{evolveNE}
\bea
\label{evloveN}
\frac{{\rm d}N}{{\rm d}t} &=& 
\left[ -G_N + T_N + c_{11}N N^{\dagger} + c_{12}E E^{\dagger} \right] N, \\
\label{evolveE}
\frac{{\rm d}E}{{\rm d}t} &=& 
\left[ -G_E + T_E + c_{21}E E^{\dagger} + c_{22}N N^{\dagger} \right] E,
\eea
\end{mathletters}
where $t \equiv (1/16\pi^2) \ln (\mu/M_Z)$, and $G_N$, $G_E$, $T_N$,
$T_E$ and $c_{ij}$ are listed in Table I with $U$ and $D$ being the
$3\times3$ up-type and down-type Yukawa coupling matrices,
respectively\footnote{The trace part of radiative corrections comes
from the fermion loops on the Higgs fields where quark contributions
enter.}.
\begin{table}[ht]
\begin{tabular}{c|ccc}
Models & SM & THDM & MSSM \\
$G_N$ & $\frac{9}{20} g_1^2 + \frac94 g_2^2$ & 
$\frac{9}{20} g_1^2 + \frac94 g_2^2$ &
$\frac35 g_1^2 + 3 g_2^2$ \\
$G_E$ & $\frac94 g_1^2 + \frac94 g_2^2$ & 
$\frac94 g_1^2 + \frac94 g_2^2$ & 
$\frac95 g_1^2 + 3 g_2^2$ \\
$T_N$ & 
${\rm Tr}\left[ 3U U^{\dagger} + 3D D^{\dagger} 
               + N N^{\dagger} + E E^{\dagger} \right]$ & 
${\rm Tr}\left[ 3U U^{\dagger} + N N^{\dagger} \right]$ & 
${\rm Tr}\left[ 3U U^{\dagger} + N N^{\dagger} \right]$ \\
$T_E$ & 
${\rm Tr}\left[ 3U U^{\dagger} + 3D D^{\dagger} 
               + N N^{\dagger} + E E^{\dagger} \right]$ & 
${\rm Tr}\left[ 3D D^{\dagger} + E E^{\dagger} \right]$ & 
${\rm Tr}\left[ 3D D^{\dagger} + E E^{\dagger} \right]$ \\
$c_{11}$ & $\frac32$ & $\frac32$ & $3$ \\
$c_{12}$ & $-\frac32$ & $\frac12$ & $1$ \\
$c_{21}$ & $\frac32$ & $\frac32$ & $3$ \\
$c_{22}$ & $-\frac32$ & $\frac12$ & $1$ \\
\end{tabular}
\vspace{6pt}
\caption{}
\end{table}
Using Eqs.~(\ref{detC}) and (\ref{evolveNE}), one obtains
\be
\frac{{\rm d}C}{{\rm d}t}=\{C,A\},
\ee
where $A = -(G_N+G_E) + (T_N+T_E) + (2 c_{11} + c_{12}) N^{\dagger} N
+ (2 c_{21} + c_{22}) E^{\dagger} E$.  Since
\bean
\frac{{\rm d} \ln {\rm det} C}{{\rm d} t} = 2\,{\rm Tr} A,
\eean
we obtain a simple evolution equation
\be
\frac{({\rm det} C)_{\mu}}{({\rm det} C)_{M_Z}} = 
{\rm exp}\left[ \int_0^{\tau} (2\,{\rm Tr} A) {{\rm d}t} \right].
\ee
Assuming a hierarchy structure in the Yukawa couplings, one can neglect
contributions from $D$, $N$ and $E$ as unimportant because at the
leading order of the perturbative correction the top quark loop in the
Higgs field renormalization dominates.  Some comments are in order:
\begin{itemize}

\item Assuming that all contributions to the Yukawa coupling evolution
are due to a heavy top quark, we find that all lepton Yukawa couplings
scale in the same way \cite{L1987}.  Therefore, the ratio $(\Delta
M^2)_{\mu}/(\Delta M^2)_{M_Z}$ is independent of the lepton masses,
where $\Delta M^2 \equiv (m_{\tau}^2-m_{\mu}^2) (m_{\mu}^2-m_e^2)
(m_e^2-m_{\tau}^2) (m_{\nt}^2-m_{\nm}^2) (m_{\nm}^2-m_{\ne}^2)
(m_{\ne}^2-m_{\nt}^2)$.

\item Most of the running of ${\rm det}\,C$ is due to the running of
masses (instead of that of the Jarlskog invariant $J$), which is
mainly a result of the dominant top quark Yukawa coupling.  By
comparing the running of ${\rm det}\,C$ and $\Delta M^2$, we find
that, as expected, $J$ does not change very much over the energy scale
range considered.

\end{itemize}

When running ${\rm det}\,C$ in the three models considered, the vacuum
expectation energy of the Higgs field is taken to be $v = 174$ GeV in
the SM and $v = \sqrt{v_u^2 + v_d^2}$ with $\tan \beta \equiv v_u/v_d
= 10$ \footnote{The value of $\tan \beta$ cannot be too large if the
top quark effect is considered to be dominant.  If $\tan \beta$ is
very large, bottom quark contributions to the evolution should be
taken into account too.} in both the THDM and MSSM.  With $y_t$ being
the Yukawa coupling of the top quark, one has \cite{L1987,CEL1974}
\bea
\label{trA}
{\rm Tr}A \simeq 
\left\{
\bary {l}
-3 \left(\frac{54}{20} g_1^2 + \frac92 g_2^2\right) + 18 y_t^2 
\;\; {\rm (SM)}, \\
-3 \left(\frac{54}{20} g_1^2 + \frac92 g_2^2\right) + 9 y_t^2 
\;\; {\rm (THDM)}, \\
-3 \left(\frac{12}{5} g_1^2 + 6 g_2^2\right) + 9 y_t^2 
\;\; {\rm (MSSM)}.
\eary
\right.
\eea
As one can see from Eqs.~(\ref{trA}), the evolution of ${\rm det} C$
for leptons differs from that for quarks because it does not involve
the QCD coupling, which significantly drags down the running quark
masses.

\begin{figure}[ht]
\centerline{\epsfysize=8truecm  \epsfbox{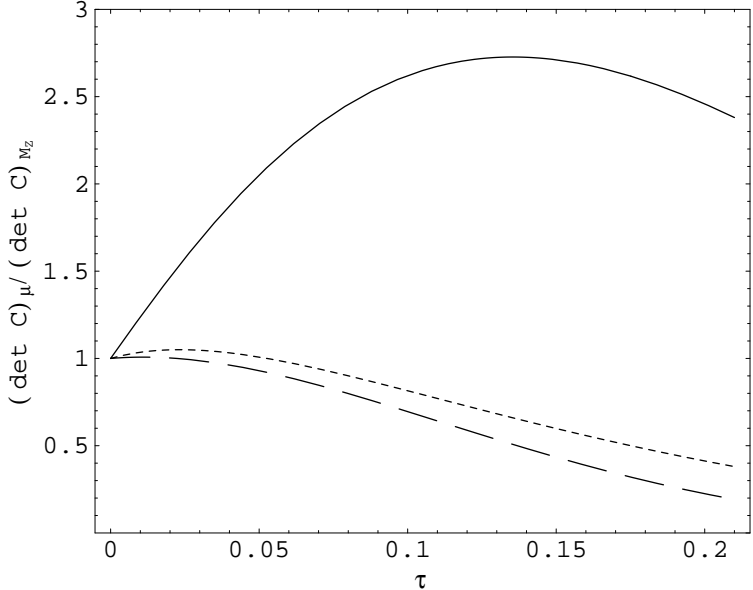} }
\vspace{10pt}
\tighten{
\caption[]{The evolution of $({\rm det}\,C)_{\mu}/({\rm
det}\,C)_{M_Z}$ with the scale $\mu$.  We assume Dirac masses for
neutrinos.  The solid curve is for the Standard Model, the dotted
curve is for the two Higgs model, and the dashed curve is for the
minimal supersymmetric model.}}
\end{figure}

The evolution of the ratio $({\rm det}\,C)_{\mu}/({\rm det}\,C)_{M_Z}$
as a function of the energy scale $\mu$ for Dirac neutrinos is shown
in FIG. 1.  It is seen that ${\rm det}\,C$ has a very distinct
behavior in the SM that is different from that in the THDM and MSSM.
It is the result of a stronger dependence on $y_t$ for ${\rm Tr}A$ in
the SM in Eq.~(\ref{trA}) and the evolution of $y_t$ itself.  A convex
curve for the SM follows from $y_t$ becoming smaller at larger scales,
thus decreasing ${\rm Tr}A$.  The minor difference between the THDM
and MSSM is due to the differences between the coefficients of $g_1$
and $g_2$.  Therefore, the dependence on $y_t$ is the dominant factor
in determining the evolution of $({\rm det}\,C)_{\mu}/({\rm
det}\,C)_{M_Z}$.  Within the SM, ${\rm det}\,C$ becomes larger as the
energy scale goes up; whereas in both the THDM and MSSM, ${\rm
det}\,C$ evolves to a smaller value.  At the GUT scale, ${\rm det}\,C$
in the SM is $\sim 6$ times bigger than in the THDM and $\sim 12$
times than in the MSSM.  The curves for the scale dependence of
$\Delta M^2$ are almost identical to those for ${\rm det}\,C$.
Numerically, $({\rm det}\,C,\Delta M^2)$ at the GUT scale is
$(2.42,2.42)$, $(0.40,0.39)$, and $(0.21,0.19)$ in the SM, THDM and
MSSM, respectively.

If a smaller value of $\tan \beta$ is used, the ratio $({\rm
det}\,C)_{\mu}/({\rm det}\,C)_{M_Z}$ in the THDM and MSSM will behave
more like that in the SM.  This is because in this case $y_t$ is
larger at the electroweak scale compared to the case with a larger
$\tan \beta$.  One should also note that in some circumstances the
evolution of $y_t$ blows up because within the dominant top Yukawa
coupling assumption,
\bea
\label{yt}
y_t(\tau) = \frac{y_t(0) E(\tau)}
                 {1 - c \, y_t(0) \int_0^{\tau} E(t) {\rm d}t},
\eea
where $E(\tau) = \exp \left[ - \int_0^{\tau} a_i \, g_i^2 {\rm d}t
\right]$ and $c>0$.  The numerical values of $a_i$ with $i=1,2,3$ and
$c$ depend upon the model.  Eq.~(\ref{yt}) tells us that either one
has to include other fermion contributions or that higher order
perturbative corrections \cite{JM1984} should be considered when the
denominator gets close to zero.

The other interesting case occurs when the neutrino masses are
acquired through the seesaw mechanism.  In this case, we have the
effective interaction $N \, {\overline \psi_{\nu_L}} \, \psi_{\nu_L}
\, H \, H$.  This will introduce two extra loop contributions on the
Higgs legs compared to the previous case.  Therefore, in the seesaw
model both $G_N$ and $T_N$ are twice as big as for Dirac neutrinos.

\begin{figure}[ht]
\centerline{\epsfysize=8truecm  \epsfbox{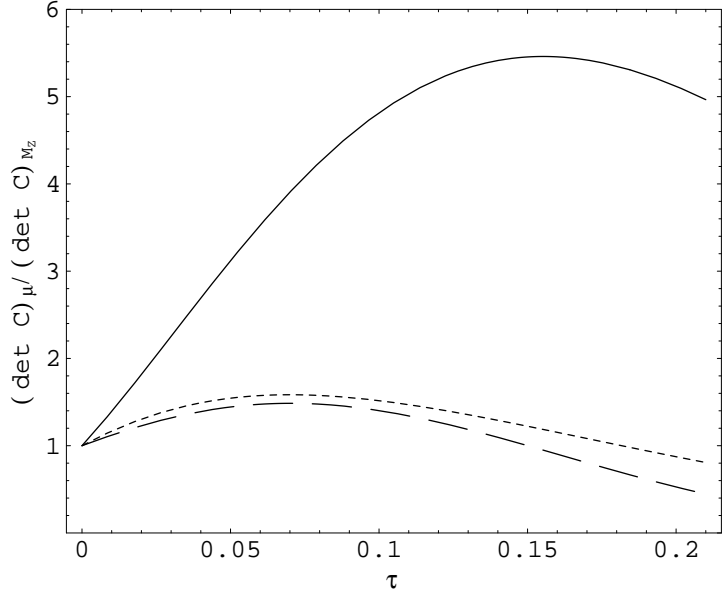} }
\vspace{10pt}
\tighten{
\caption[]{The evolution of $({\rm det}\,C)_{\mu}/({\rm
det}\,C)_{M_Z}$ with the scale $\mu$.  The labeling convention is the
same as FIG.~1, but neutrino masses are generated through the See-Saw
mechanism.}}
\end{figure}

We see in Fig.~2 that the quantity ${\rm det}\,C$ at the GUT scale is
roughly twice as big as in Fig.~1.  $({\rm det}\,C,\Delta M^2)$ at the
GUT scale is $(5.05,5.05)$, $(0.84,0.79)$, and $(0.49,0.42)$ in the
SM, THDM and MSSM, respectively.

Regardless of their different behavior, we see that for all models
considered the evolution of ${\rm det}\,C$ is, unlike the quark sector
\cite{ADG1986}, of order $10^{-1}$ to $10$ in going from the
electroweak scale to the GUT scale.  The actual factors that enter a
$CP$-violating quantity are not necessary only ${\rm det}\,C$, but may
involve other mass scales relevant to the process.  The $CP$-violating
mechanism responsible for leptogenesis may have a different origin.
As long as the physics the gives rise to lepton asymmetry effects at a
higher energy scale is in the same multiplet of the Standard Model
particles within the unified theory, as an effective theory, the
quantity ${\rm det}\,C$ will still be a good indicator of how much
$CP$ nonconservation would be at higher energy scales, though the
exact relation depends upon models.

As delineated in Ref.~\cite{DLRW2000} with a toy model containing two
very heavy SU(2)-doublet scalars, if the neutrinos acquire small Dirac
masses, the $CP$-violating processes due to the heavy scalars at the
large energy scale can produce a neutrino asymmetry.  Both the heavy
scalars and the light Dirac neutrinos are out of equilibrium at the
weak scale where sphalerons come into play to produce baryon and
lepton number violation.  Similarly, if the neutrino masses are
generated through the seesaw mechanism with very heavy ($\sim 10^{16}$
GeV) singlet right-handed neutrinos, then both the out-of-equilibrium
and $CP$ violation conditions can be satisfied around the GUT scale.
One then can have lepton asymmetry leading to baryogenesis
\cite{FY1986}.

We conclude that hierarchical neutrino masses suggest that the
possible leptonic $CP$ violation strength at the GUT scale differs
from that at the electroweak scale by no more than one order of
magnitude.  Since this quantity is reparametrization invariant, models
of grand unified theories should give consistent predictions for it.
The fact that the measure of the leptonic $CP$ nonconservation does
not have a significant decrease at higher scales may help leptogenesis
in the early Universe.

The author would like to thank F.J. Gilman, I.Z. Rothstein, and
L. Wolfenstein for useful discussions and A.K. Leibovich for comments.
This work is supported by the Department of Energy under Grant
No.~DE-FG02-91ER40682.

\end{document}